\documentclass[11pt]{article}

\usepackage[percent]{overpic}
\usepackage{float}
\usepackage{pgfplots}
\usepackage{placeins}
\usepackage{amsmath}
\usepackage{amsthm}
\usepackage{color}
\usepackage{amssymb}
\usepackage{mathtools}
\usepackage{subfigure}
\usepackage{multirow}
\usepackage{epsfig}
\usepackage{listings}
\usepackage{enumitem}
\usepackage{rotating,tabularx}
\usepackage{graphicx}
\usepackage{graphics}
\usepackage{epstopdf}
\usepackage{longtable}
\usepackage[pdftex]{hyperref}
\usepackage{epigraph}
\usepackage{xspace}
\usepackage{amsfonts}
\usepackage{eurosym}
\usepackage{ulem}
\usepackage{footmisc}
\usepackage{comment}
\usepackage{setspace}
\usepackage{geometry}
\usepackage{caption}
\usepackage{pdflscape}
\usepackage{array}
\usepackage[round]{natbib}
\usepackage{booktabs}
\usepackage{dcolumn}
\usepackage{mathrsfs}

\usepackage{enumitem}
\usepackage{tikz}
\usetikzlibrary{decorations.pathreplacing}

\usepackage{xcolor}
\hypersetup{
colorlinks,
linkcolor={blue!50!black},
citecolor={blue!50!black},
urlcolor={blue!50!black}}

\newcommand{\bc}{\begin{center}}
\newcommand{\ec}{\end{center}}

\newcolumntype{d}[1]{D{.}{.}{#1}} 

\newcolumntype{L}[1]{>{\raggedright\let\newline\\arraybackslash\hspace{0pt}}m{#1}}
\newcolumntype{C}[1]{>{\centering\let\newline\\arraybackslash\hspace{0pt}}m{#1}}
\newcolumntype{R}[1]{>{\raggedleft\let\newline\\arraybackslash\hspace{0pt}}m{#1}}

\newcommand{\elabel}[1]{\label{eq:#1}}
\newcommand{\eref}[1]{(\ref{eq:#1})}
\newcommand{\Eref}[1]{(\ref{eq:#1})}

\newcommand{\seclabel}[1]{\label{sec:#1}}
\newcommand{\Secref}[1]{Section~\ref{sec:#1}}
\newcommand{\secref}[1]{Section~\ref{sec:#1}}

\newcommand{\appref}[1]{Appendix~\ref{app:#1}}

\newcommand{\ie}{{\it i.e.}\xspace}
\newcommand{\eg}{{\it e.g.}\xspace}

\newcommand{\ave}[1]{\left\langle#1 \right\rangle}
\newcommand{\aveN}[1]{\left\langle#1 \right\rangle_N}

\newcommand{\flabel}[1]{\label{fig:#1}}
\newcommand{\fref}[1]{Figure~\ref{fig:#1}}
\newcommand{\Fref}[1]{Figure~\ref{fig:#1}}

\newcommand{\be}{\begin{equation}}
\newcommand{\ee}{\end{equation}}
\newcommand{\bea}{\begin{eqnarray}}
\newcommand{\eea}{\end{eqnarray}}

\newcommand{\bi}{\begin{itemize}}
\newcommand{\ei}{\end{itemize}}

\newcommand{\Dt}{\Delta t}
\newcommand{\Dx}{\Delta x}

\newcommand{\Ito}{It\^{o}\xspace}

\newcommand{\D}{\Delta}
\newcommand{\dx}{\delta x}
\newcommand{\dt}{\delta t}

\newcommand{\N}{\mathcal{N}}

\newcommand{\gt}{\bar{g}}
\newcommand{\ga}{g_{\ave{}}}
\newcommand{\gdem}{\bar{g}_{N}}
\newcommand{\gplu}{g_{\ave{}_N}}

\definecolor{col1}{rgb}{0.96, 0.94, 0.93} 

\definecolor{col2}{rgb}{0, 0, 0} 

\numberwithin{equation}{section}
\DeclareMathOperator\erf{erf}

\begin{document}

\begin{titlepage}
\title{The Two Growth Rates of the Economy}
\author{Alexander Adamou\footnote{London Mathematical Laboratory,~\url{a.adamou@lml.org.uk}} \and Yonatan Berman\footnote{London Mathematical Laboratory and Stone Center on Socio-Economic Inequality,~\url{y.berman@lml.org.uk}} \and Ole Peters\footnote{London Mathematical Laboratory and Santa Fe Institute,~\url{o.peters@lml.org.uk}}\,\, \thanks{The code to generate all figures is available at \url{https://bit.ly/DDPcodes}.}}
\date{\today}
\maketitle


\begin{abstract}
\noindent Economic growth is measured as the rate of relative change in gross domestic product (GDP) per capita. Yet, when incomes follow random multiplicative growth, the ensemble-average (GDP per capita) growth rate is higher than the time-average growth rate achieved by each individual in the long run. This mathematical fact is the starting point of ergodicity economics. Using the atypically high ensemble-average growth rate as the principal growth measure creates an incomplete picture. Policymaking would be better informed by reporting both ensemble-average and time-average growth rates. We analyse rigorously these growth rates and describe their evolution in the United States and France over the last fifty years. The difference between the two growth rates gives rise to a natural measure of income inequality, equal to the mean logarithmic deviation. Despite being estimated as the average of individual income growth rates, the time-average growth rate is independent of income mobility.

\end{abstract}
\setcounter{page}{0}
\thispagestyle{empty}
\end{titlepage}
\pagebreak \newpage

\section{Introduction}
\seclabel{Introduction}

Ergodicity economics is a perspective that emerges when we ask explicitly whether stochastic processes used as economic models are ergodic \citep{Peters2019b}. If a stochastic process is ergodic, its ensemble average\footnote{The ensemble average of a stochastic process $Y(t)$ is the large-$N$ limit of the average value over $N$ realisations. This limit is also called the expected value, or expectation value, or first moment, or mean, denoted $\ave{Y(t)}\equiv\lim_{N\to\infty}\frac{1}{N}\sum_{i=1}^N y_i(t)$.} is identical to its time average.\footnote{The time average of a stochastic process $Y(t)$ is the long-time limit of the average value along a trajectory $y(t)$, denoted $\bar{Y}\equiv\lim_{T\to\infty}\frac{1}{T}\int_{t}^T y(s)\,ds$.}
An important counterexample is multiplicative growth with random shocks, a widely used model of growth in economics and elsewhere. Here, the ensemble average grows exponentially at a deterministic rate, which we call the {\it ensemble-average growth rate}. In the long-time limit, any individual trajectory also grows exponentially at a deterministic rate, which we label the {\it time-average growth rate}. Crucially, these two rates differ: the ensemble-average growth rate is greater than the time-average growth rate.

In this paper we explore what ergodicity economics tells us about measures of national economic performance. We find that the two growth rates described above correspond to two types of economic growth that are discussed in various guises in the literature.

The ensemble-average growth rate is closely related to the most quoted measure of national economic performance, namely to the growth rate of gross domestic product (GDP) per capita. GDP is the total income generated by the economic activity of a national population, so the growth rate of GDP per capita is simply the growth rate of average income. For large populations, this coincides with the ensemble-average growth rate. The GDP per capita growth rate is a so-called \textit{plutocratic} growth measure in the following sense: it is approximately a weighted average of individual income growth rates, where the weight assigned to each individual is proportional to their income.\footnote{This nomenclature is standard. For example, \citet{Milanovic2005} uses the adjective ``plutocratic'' to describe statistics with this property. We make the term mathematically precise in \secref{theory}.}

The time-average growth rate is representative of individual growth, being the growth rate realised with certainty by each income in the long-time limit. Unlike the ensemble-average growth rate, it is rarely measured or stated. The time-average growth rate can be estimated as the average of the individual income growth rates. This makes it a \textit{democratic} growth measure, because each person is assigned an equal weight in the average. To keep the symmetry between ensemble averages and time averages, and to provide a handy nomenclature, we introduce the \textit{democratic domestic product} (DDP) per capita as the quantity that grows at the time-average growth rate.\footnote{Existing names for the same concept are the ``people's growth rate'' \citep{Milanovic2005,SaezZucman2019b} and the ``population-weighted growth rate'' \citep{Milanovic2005}. As far as we know, the distinction between the two growth rates was first drawn by \citet{CheneryETAL1974}, when discussing different ways to weight quintile average income growth rates to obtain a single indicator of economic performance.}

The ensemble-average (GDP per capita) growth rate may measure changes in a nation's total productive capacity, but it is an unsuitable measure of the typical economic experience of the nation's citizens. Conversely, the time-average (DDP per capita) growth rate may reflect how citizens experience the economy, but it is not a good measure of what's happening to total productive capacity.

The tension between these two growth rates gives rise to a natural measure of inequality \citep{AdamouPeters2016}. If the ensemble-average growth rate exceeds the time-average growth rate, \ie if GDP per capita grows faster than DDP per capita, then inequality is increasing. If the converse is true, then inequality is decreasing. This suggests defining inequality simply as the quantity which grows at the difference between the two rates, ensemble-average minus time-average. We call this inequality measure the \textit{ergodicity gap}. Under multiplicative growth, it corresponds to an established inequality metric: the \textit{mean logarithmic deviation} (MLD), also known as the second index of \citet{Theil1967}. At perfect equality, the ergodicity gap vanishes, and GDP per capita and DDP per capita are equal.  

Computing the time-average (DDP per capita) growth rate as the average of individual growth rates requires, \textit{prima facie}, hard-to-obtain panel data. However, we show that this growth rate and, by extension, the ergodicity gap (MLD) depend only on the initial and terminal income distributions. They do not depend on the reordering of individuals that occurs in-between, \ie they are independent of rank mobility. This has the practical benefit of allowing their estimation using cross-sectional income data.

Our contribution is twofold. First, we relate the ensemble-average and time-average growth rates of ergodicity economics to the two types of growth measure discussed in the literature. We use them to define a fundamental measure of inequality. To better inform policymaking, we propose that national statistical authorities compile and report all three measures. Second, we illustrate potential uses for these statistics by studying their strikingly different evolutions in the United States and France over the past fifty years.

The existence of the two growth measures raises the possibility of political disconnect. To a central planner, say the government in Washington, DC, the most visible figure is GDP. It is an accurate measure of the nation's productive capacity and central planners are right to be interested in it. But the situation may arise, and data suggest it has arisen, where ensemble-average growth is consistently faster than time-average growth. Central planners may see `the economy' doing well while, on the ground, individuals are doing less well. If allowed to persist, this situation can lead to a disconnect between the central government and the typical citizen. Citizens may feel left behind, excluded from economic prosperity, or even lose faith in their institutions and press, who promulgate statistics that paint a picture of prosperity unrecognisable to most.

\section{The two growth rates of the economy}
\seclabel{theory}

Ergodicity economics places great emphasis on the functional form of the growth rate \citep{PetersAdamou2018a,Peters2019b}. In simple terms, a growth rate extracts the scale parameter of time in a growth function. Different types of growth correspond to different growth functions and, therefore, to different growth rates. For example, a quantity that grows linearly in time according to the growth function $y=rt$, has its growth rate extracted by the operation $\mathrm{d} y/\mathrm{d} t=r$; while under an exponential growth function, $y=\exp(rt)$, the growth rate is extracted by $\mathrm{d}\ln y/\mathrm{d} t=r$. In both cases, a constant is obtained from a time-varying function. At a deeper level, and generalising to growth with fluctuations, the role of the growth rate is to extract a stationary mathematical object -- specifically an ergodic observable -- from a model of a growing observable. This is achieved by identifying an ergodicity transformation, which maps the observable of interest (here, income) to an observable whose rate of change is ergodic. Finding this transformation is often a powerful step in ergodicity economics, as we shall see in \secref{time_average}. In the model of income we use here, we show that the transformation is simple (the logarithm) and we do not dwell on its deeper meaning.

\subsection{Random multiplicative growth}
\seclabel{rand_mult_grow}

In random multiplicative growth, income obeys Gibrat's law of proportionate effect \citep{Gibrat1931}, \ie incomes are multiplied by random variables at successive time steps. Variants of this model have been studied since the 1930s. We recommend \citet[Ch.~11]{AitchisonBrown1957} for a review of early research and \citet{GabaixETAL2016} for up-to-date coverage. We consider random multiplicative growth a null model of income, which captures two dominant effects: that incomes fluctuate; and that they do so in proportion to themselves.\footnote{Changes in income, such pay rises and cuts, are typically expressed as percentages per year, as are changes in regular outgoings, like housing and transport costs. We do not suggest this captures all aspects of reality that determine people's incomes, just the most important ones. We note that predictions about the distribution of income under random multiplicative growth can be tested empirically \citep{GuvenenETAL2017} and that extensions to general growth processes are available \citep{PetersAdamou2018a}.}

We consider a model population of $N$ people, labelled $i=1\dots N$, whose incomes, $x_i(t)$, are functions of time. Income is measured in units of currency per unit time, \eg dollars per year, and is always positive. Income at one observation time is the income at the previous observation time multiplied by a random variable,
\be
x(t+\dt) = x(t)(1+\epsilon)\,,
\elabel{disc_mult}
\ee
where $\dt$ is the time between observations and $\epsilon$ is the random variable. \Eref{disc_mult} defines a stochastic process which generates trajectories of income,
\be
x(t+T\dt) = x(t)\prod_{\tau=1}^T(1+\epsilon_\tau)\,,
\elabel{disc_prod}
\ee
where $\epsilon_\tau$ is the realisation of $\epsilon$ corresponding to the $\tau^\text{th}$ observation. We assume $\epsilon$ is stationary, \ie that its distribution does not change in time, and that realisations at different times $\tau$ are independent. Note that this makes the stochastic process $\epsilon_\tau$ ergodic: the time average and ensemble average of $\epsilon_\tau$ are trivially identical.

The rate of change of income over the single time step in \eref{disc_mult} is $\dx/\dt = x(t)\epsilon/\dt$. This is not ergodic because it depends on $x(t)$.\footnote{The ensemble-average is $\ave{\dx/\dt}=x(t)\ave{\epsilon}/\dt$. The time-average is the average value of $\dx/\dt$ along an income trajectory, which will diverge or be zero depending on whether $x(t)$ grows or decays.} Therefore, we introduce an ergodicity transformation, \ie a function of $x_i(t)$ whose rate of change is ergodic. We do this by requiring the change in the function to be independent of $x(t)$. Taking the logarithm of \eref{disc_mult} turns the product of income and income multiplier into a sum, which we rearrange as
\be
\ln x(t+\dt) - \ln x(t) = \ln (1+\epsilon)\,.
\elabel{disc_add}
\ee
This is the desired result: the left-hand side is the additive change of transformed income; the right-hand side is a stationary random variable with no $x$-dependence. The rate of change, \mbox{$g = \delta\ln x(t)/\dt$}, is now ergodic, $\ave{g}=\bar{g}$, and we call it the ergodic growth rate.\footnote{The ensemble-average is
\be
\ave{g} = \lim_{N\to\infty} \frac{1}{N} \sum_{i=1}^N \frac{\delta\ln x_i(t)}{\dt} = \lim_{N\to\infty} \frac{1}{N} \sum_{i=1}^N \frac{\ln(1+\epsilon_i)}{\dt}\,,
\ee
where $i$ labels different trajectories of the income process, \eref{disc_prod}, at a single point in time. The time-average is
\be
\bar{g} = \lim_{T\dt\to\infty} \frac{1}{T\dt} \sum_{\tau=0}^{T-1} \frac{\delta\ln x(t+\tau\dt)}{\dt}\,\dt = \lim_{T\to\infty} \frac{1}{T} \sum_{\tau=0}^{T-1} \frac{\ln(1+\epsilon_\tau)}{\dt}\,,
\ee
where $\tau$ labels different times of a single trajectory of the income process, \eref{disc_prod}. Since $\epsilon$ is stationary and its realisations are independent, the two averages have the same limiting value.}

Since this analysis does not depend on the length of time, $\dt$, between observations, we write the growth rate of individual income over a general time period from time $t$ to time $t+\Dt$ as
\be
g_i = \frac{\D\ln x_i(t)}{\Dt}\,,
\elabel{gi}
\ee
where $\D\ln x_i(t) = \ln x_i(t+\Dt) - \ln x_i(t)$ is the change in the logarithm of the $i^\text{th}$ income.

\subsection{Ensemble-average growth rate}
\seclabel{ensemble_average}

We denote by $\aveN{x(t)}$ the population-average income,
\be
\aveN{x(t)} \equiv \frac{1}{N}\sum_{i=1}^N x_i(t)\,,
\ee
which we can compute from observations of real incomes (or estimate from partial data). In a model of income dynamics, in which we can generate as many model incomes as we like, the ensemble-average income, $\ave{x(t)}$, is the $N\to\infty$ limit of the population-average income,
\be
\ave{x(t)} \equiv \lim_{N\to\infty} \aveN{x(t)}\,.
\ee
%

In practice, the national populations we study are large enough for differences between modelled values of $\ave{x(t)}$ and $\aveN{x(t)}$ to be negligible (see \appref{finite} for details). Therefore, we shall refer to both quantities as the \textit{ensemble-average income}, using the subscript $N$ to distinguish the finite-population statistic from that of the infinite ensemble.

The ensemble-average income, being the total personal income of a population divided by the number of people, resembles the gross domestic product (GDP) per capita. We shall make this association explicit throughout the text.

The ensemble-average income (GDP per capita) grows at the rate
\be
\ga \equiv \frac{\D\ln\ave{x(t)}}{\Dt}\,.
\elabel{ga}
\ee
or, for a finite population, at the rate
\be
\gplu \equiv \frac{\D\ln\aveN{x(t)}}{\Dt}\,.
\ee

We call this the \textit{ensemble-average growth rate} of income (GDP per capita growth rate).\footnote{What we call the ensemble-average growth rate (GDP per capita growth rate) is the growth rate of the ensemble average (GDP per capita) and, we stress, not the ensemble average of the growth rate.}

If changes in income are small, then we can approximate the ensemble-average (GDP per capita) growth rate as a weighted average of individual growth rates
\be
\gplu \approx \sum_{i=1}^N w_i g_i\,,
\footnote{For small income changes, $\D\ln x_i \approx \Dx_i/x_i$ and
\bea
\frac{\D\ln\aveN{x(t)}}{\Dt} &\approx& \frac{\D\aveN{x}/\aveN{x}}{\Dt} \\
&=& \frac{\sum_{i=1}^N \Dx_i}{\sum_{i=1}^N x_i \,\Dt} \\
&=& \sum_{i=1}^N \frac{x_i}{\sum_{i=1}^N x_i} \frac{\Dx_i/x_i}{\Dt} \\
&\approx& \sum_{i=1}^N w_i \frac{\D\ln x_i}{\Dt}\,,
\eea
where $w_i=x_i/\sum_{i=1}^N x_i$ as required.}
\ee
where each person's growth rate is weighted by their share of the total income, $w_i=x_i/\sum_{i=1}^N x_i$. The ensemble-average (GDP per capita) growth rate is, therefore, a \textit{plutocratic} measure of growth \citep{Milanovic2005}, in the sense that each person's weight in the average is proportional to their income.

\subsection{Time-average growth rate}
\seclabel{time_average}

Whereas the ensemble-average growth rate requires a model where we can average over indefinitely many individuals, the \textit{time-average growth rate} of income requires just one modelled individual but an indefinitely long income trajectory. It is defined as the $\Dt\to\infty$ limit of the individual growth rate in \eref{gi},
\be
\gt \equiv \lim_{\Dt\to\infty} \frac{\D\ln x_i(t)}{\Dt}\,.
\elabel{gt}
\ee
Conceptually, therefore, the time-average growth rate measures the typical growth of an individual, experienced in a single income trajectory unfolding over time.
 
Alternatively, $\gt$ can be expressed as a time average if we partition the period $\Dt$ into $T$ shorter periods of fixed duration $\dt$:
\be
\gt = \lim_{T\to\infty} \frac{1}{T} \sum_{\tau=0}^{T-1} \frac{\delta\ln x_i(t+\tau\dt)}{\dt}\,.
\elabel{gt_ave}
\ee
Thus we see that the two different procedures -- computing a long-time growth rate and averaging a short-time growth rate over many time steps -- are equivalent. This convenient property follows from our definition of the growth rate as the (additive) rate of change of a function of the process.

Of course, we cannot compute any infinite-time limit from data of real incomes. However, we can exploit the ergodicity of the growth rate, $\bar{g}=\ave{g}$, derived in \secref{rand_mult_grow}, to express the time-average growth rate as an ensemble average,
\be
\gt = \ave{\frac{\D\ln x(t)}{\Dt}}\,,
\ee
which for a large population we can estimate as a finite-$N$ average,
\be
\gdem \equiv \aveN{\frac{\D\ln x(t)}{\Dt}}\,.
\elabel{gt_aveN}
\ee
\Eref{gt_aveN} is our practical definition of the time-average growth rate of income.

Our ability to estimate the time-average growth rate as an ensemble average is worth pondering. At any moment in time there exists a long income trajectory that would result from experiencing the ensemble of short-time ergodic growth rates in sequence. We speculate that this momentary estimate of the long-time growth rate corresponds closely to how an economy feels to an individual. It tells the individual what he or she can expect in the long run if the balance of risks and opportunities stays the same.\footnote{Another interpretation of $\gdem$ is the growth rate a hypothetical person would experience if they spent $1/N$ of each year growing like every person in the population. Since $\gdem=\aveN{g}$, we have
\be
x(t) \exp(\gdem\Dt) = x(t) \exp \left( \sum_{i=1}^N \frac{g_i\Dt}{N} \right)\,.
\ee
This everyman growing at $\gdem$, unlike his lucky compatriot growing at $\gplu$, shares in all of the experiences of the national economy in a single year. A truly representative agent.}

The time-average growth rate is an average of individual growth rates, in which each person's growth rate has equal weight. This makes it a \textit{democratic} measure of growth, as explained in \secref{Introduction}, and we also call it the DDP per capita growth rate.

DDP per capita is the quantity, $y(t)$, that grows at the time-average growth rate, \ie $y(t+\Dt)=y(t)\exp(\gdem\Dt)$. Rearranging gives the DDP per capita as
\be
y(t) = \exp(\aveN{\ln x(t)})\,,
\ee
which is the geometric mean income and has the same units as income.

\Eref{gt_aveN} demonstrates that knowledge of individual growth rates is unnecessary for computing $\gdem$. We need only the values of $\aveN{\ln x}$ at the start and end of the observation period. This has practical importance, since reliable panel data, required for estimating individual growth rates, are harder to obtain than data on the initial and terminal distributions. It also means that $\gdem$ is independent of income mobility, \ie of the reordering that takes place between observations.

\subsection{Inequality}
\seclabel{inequality}

The ensemble-average (GDP per capita) growth rate and time-average (DDP per capita) growth rate are conceptually different. The former is the growth rate of the average (or, multiplying by $N$, the total) income in a large population. The latter is the growth rate achieved by a single member of the population over long time. Under multiplicative growth, and in general, these growth rates do not have the same value.
 
We can show this by a simple thought experiment: suppose all $N$ incomes in a population are equal at time $t$, \ie $x_i(t)=x(t)$ for all $i$, and that they evolve to different incomes, $x_i(t+\Dt)$, at time $t+\Dt$. The ensemble-average (GDP per capita) growth rate is
\be
\gplu = \frac{\ln\aveN{x(t+\Dt)} - \ln \aveN{x(t)}}{\Dt}
\elabel{ggplu}
\ee
and the time-average (DDP per capita) growth rate is
\be
\gdem = \frac{\aveN{\ln x(t+\Dt)} - \aveN{\ln x(t)}}{\Dt}\,.
\elabel{ggdem}
\ee
Since the logarithm is a concave function, the famous result of \citet{Jensen1906} implies that the difference between the growth rates is non-negative, \ie
\be
\gplu-\gdem = \frac{1}{\Dt} \left[\ln\aveN{x(t+\Dt)} - \aveN{\ln x(t+\Dt)}\right] \geq 0\,,
\elabel{gdiff}
\ee
vanishing only if the $x_i(t+\Dt)$ are equal.

This thought experiment is illuminating: we started with perfect income equality, moved to a state of income inequality, and found that $\gplu>\gdem$. In general, if the ensemble-average growth rate is higher than the time-average growth rate (GDP per capita grows faster than DDP per capita), then the current high earners grew faster than the low earners over the observation window, and inequality increased. If the converse is true, then inequality decreased.

In this picture, a natural measure of inequality (of which myriad measures exist in the literature) is the quantity which grows at the difference between the two growth rates \citep{AdamouPeters2016}. We call this measure \textit{the ergodicity gap} and denote it by $J(t)$ to highlight its connection to \citet{Jensen1906}. We have
\be
\frac{\D J}{\Dt} = \gplu - \gdem = \frac{\D\ln\aveN{x}}{\Dt} - \frac{\D\aveN{\ln x}}{\Dt}\,,
\ee
which, integrated over time, gives the ergodicity gap as
\be
J(t) = \ln\aveN{x(t)} - \aveN{\ln x(t)}\,,
\ee
up to an additive constant. $J(t)$ is an established income inequality measure, known as the \textit{mean logarithmic deviation} (MLD) and also as the second Theil index \citep{Theil1967}.\footnote{In general, the form of the ergodicity gap, like the form of the ergodic growth rate, depends on the transformation required to make income changes ergodic. Here, under multiplicative dynamics, this transformation is logarithmic.}

\subsection{Example: geometric Brownian motion}
\seclabel{example}
The results above are valid for any large population in which incomes are assumed to follow random multiplicative growth. Here we study a simple example of such dynamics: geometric Brownian motion (GBM). We do this to show how the quantities we have defined -- the ensemble-average (GDP per capita) growth rate, the time-average (DDP per capita) growth rate, and the ergodicity gap (MLD) -- behave in an analytically solved case.

GBM is a continuous-time model of income in which relative changes are normally distributed. Each income $x_i(t)$ obeys the \Ito stochastic differential equation,
\be
dx_i = x_i(\mu dt + \sigma dW_i)\,,
\elabel{gbm}
\ee
where $\mu$ and $\sigma$ are the drift and volatility parameters, $dx_i$ is the infinitesimal change in income, $dt$ is the infinitesimal change in time, and $dW_i\sim\mathcal{N}(0,dt)$ is the infinitesimal increment in a Wiener process. Because the model is well known, we only list the key model properties here, leaving derivations to \appref{model}.

Under GBM, the ensemble-average (GDP per capita) growth rate of income is
\be
\ga = \frac{\D\ln\ave{x}}{\Dt} = \mu\,,
\ee
and the time average growth rate (DDP per capita growth rate) of income is
\be
\gt = \frac{\D\ave{\ln x}}{\Dt} = \mu-\frac{\sigma^2}{2}\,.
\ee

Thus, the ensemble-average income (GDP per capita) grows faster than a single income trajectory in the long-time limit. This serves as a concrete example of the more abstract result in \eref{gdiff} and it has an important corollary: using the ensemble-average (GDP per capita) growth rate as the principal growth measure can create a misleading impression. In GBM, it is a systematic overestimate of how individual citizens experience the economy, as captured by the time-average (DDP per capita) growth rate.

\Fref{gbm_density_contour} shows the distribution of income and growth at the ensemble-average (GDP per capita) and time-average (DDP per capita) growth rates under GBM. The difference, both conceptual and quantitative, between these growth rates is the central observation in this study.

\begin{figure}[!htb]
\centering
\includegraphics[width=1.0\textwidth]{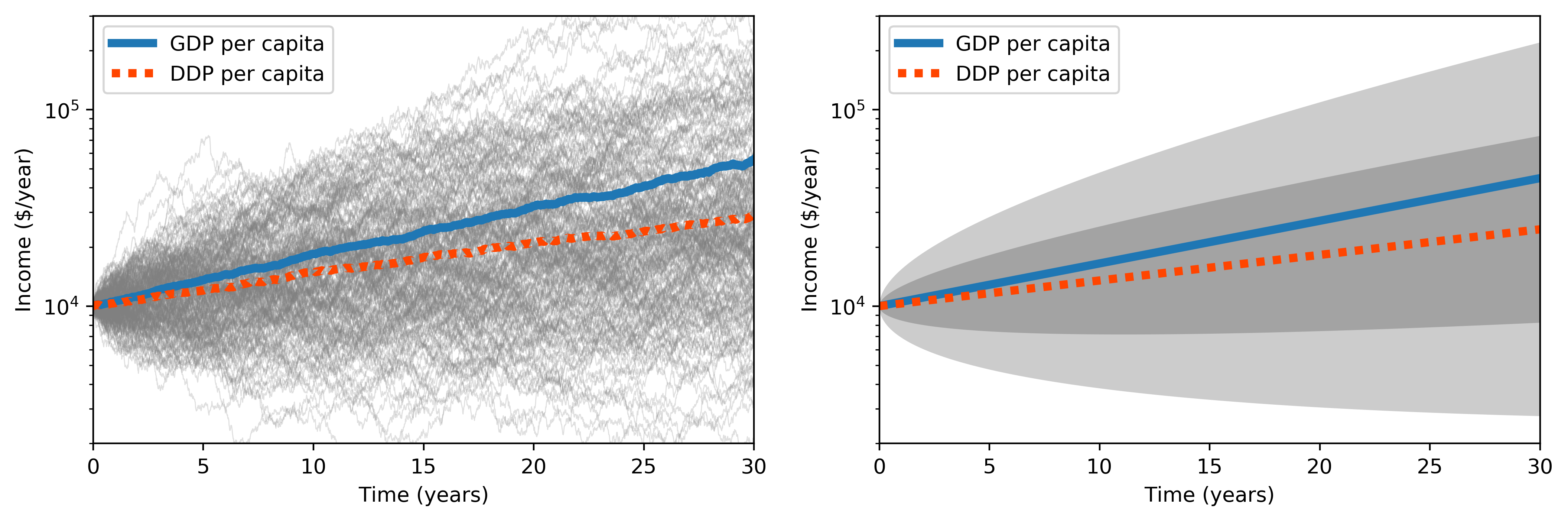}
\caption{The distribution of income in GBM. The blue line is the ensemble-average income (GDP per capita), while the red line is the DDP per capita, which grows at the time-average growth rate. In the left panel, the thin dark lines are 200 simulated individual income trajectories. In the right panel the shaded areas represent one (darker shade) and two (lighter shade) standard deviations from the DDP per capita. The parameters used are $\mu = 0.05\text{ year}^{-1}$ and $\sigma = 0.2\text{ year}^{-\frac{1}{2}}$. We assume the initial income is $\$10^4/\text{year}$.}
\flabel{gbm_density_contour}
\end{figure}

The ergodicity gap (MLD), $J(t)$, grows at the difference between the ensemble-average (GDP per capita) and time-average (DDP per capita) growth rates. In the GBM model, this difference is
\be
\frac{dJ}{dt} = \ga - \gt = \frac{\sigma^2}{2}\,,
\ee
such that
\be
J(t) = \frac{\sigma^2 t}{2}\,,
\ee
up to an additive constant. As the distribution of income broadens with time, so income inequality -- as measured by the ergodicity gap (MLD) -- increases without bound.

GBM is a model in which members of an ensemble grow their incomes exponentially and randomly, without ever interacting. In particular, there is no reallocation of income between members, as might happen in a real economy with social institutions \citep{BermanPetersAdamou2019}. It paints a stark picture in which the aggregate income grows systematically faster than the typical income, an unintuitive result made possible only by ever-increasing inequality. Such phenomena are inherently interesting to economists, policymakers, and indeed citizens. They become easier to detect when the ensemble-average (GDP per capita) growth rate, time-average (DDP per capita) growth rate, and ergodicity gap (MLD) are reported.

\FloatBarrier

\section{Growth rates in the United States and France}
\seclabel{data}

To demonstrate the differences between $\gdem$ and $\gplu$, we use income data from the United States and France over the last 50 years.\footnote{$\gplu$ is easily estimated by using the average national income for $\ave{x}_N$ in \eref{ggplu}. Calculating $\gdem$ using \eref{ggdem} requires estimating $\aveN{\ln x}$, which is done using income quantile data (see \appref{quantile}).} \fref{ddp_us_fr_trunc} presents the evolution of $\gdem$ and $\gplu$ in these two countries. It also shows the evolution of GDP per capita and DDP per capita.

\begin{figure}[!htb]
\centering
\includegraphics[width=1.0\textwidth]{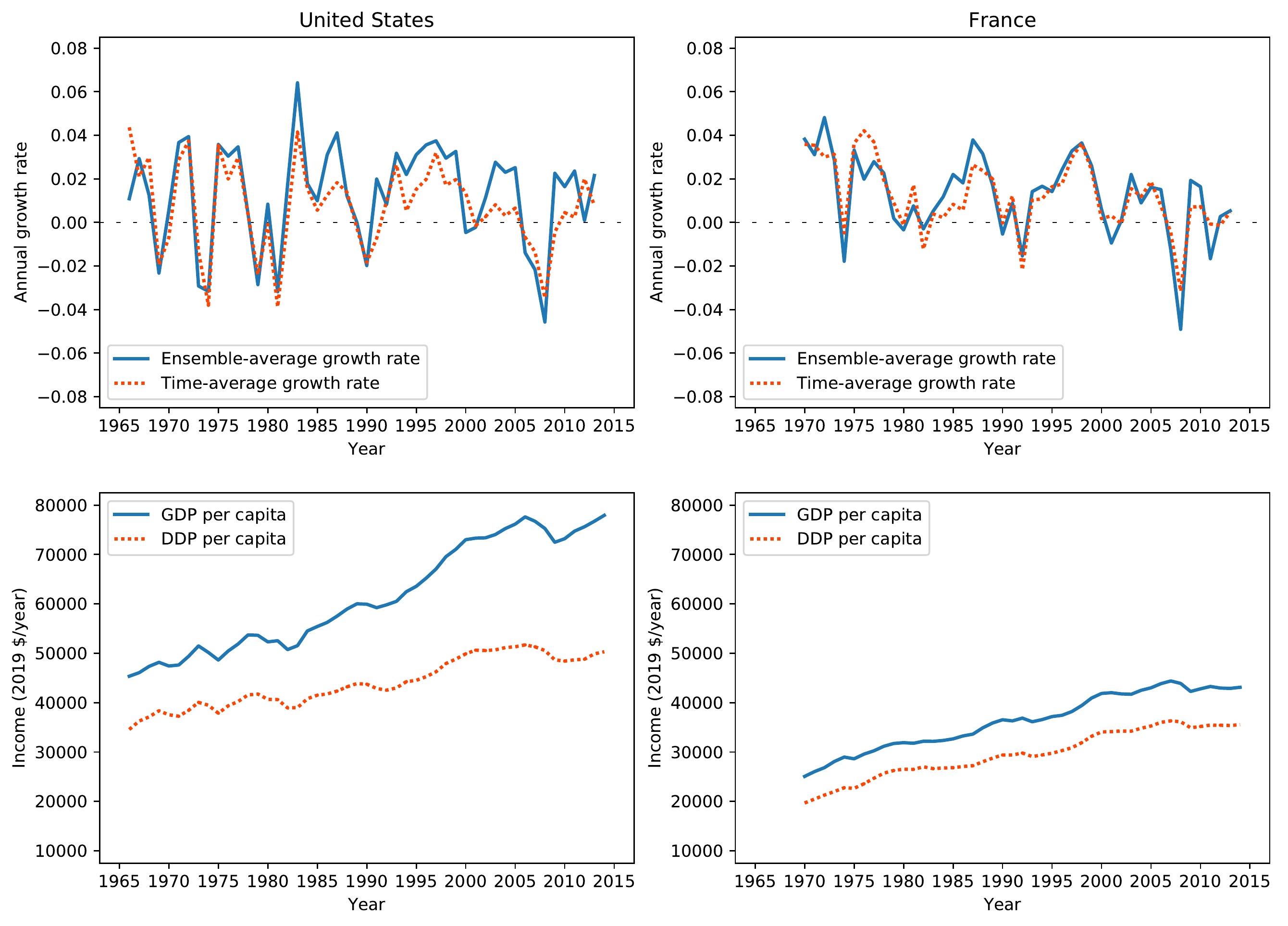}
\caption{The evolution of the two growth rates of the economy. Top: the ensemble-average (blue) and the time-average (red) growth rates in the United States (left) and in France (right). Bottom: the evolution of ensemble-average income (GDP per capita) and DDP per capita in the United States (left) and in France (right). All data were taken from \citet{WID2020}. We note that the time-average growth rate is estimated by considering the changes in the geometric mean income each year and not by averaging over all individual growth rates (see \appref{finite} for more details). We also truncated the bottom decile of the distribution (see \appref{trunc} for more details).}
\flabel{ddp_us_fr_trunc}
\end{figure}

In the United States, ensemble-average income growth (GDP per capita growth) and time-average income growth (DDP per capita growth) were similar between the mid-1960s and the early 1980s. Between 1966 and 1980, the ensemble-average income increased by 13\% and the time-average income increased by 16\%. Since 1980, there has been a clear divergence. Ensemble-average growth totalled +40\% between 1980 and 2014, whereas time-average growth totalled +20\%. This makes ensemble-average growth (meaning GDP per capita growth) a misleading quantity to follow in the United States, if the wellbeing of individuals is of interest.

In France, by contrast, the two growth rates were much closer. The ensemble-average (GDP per capita) growth rate does not create a misleading picture of the typical economic experience of France's citizens and it closely resembles the time-average (DDP per capita) growth rate. This means that a multiplicative growth model with random shocks may be inappropriate in this case. Effects that lead to divergence of incomes may be weaker here, or effects that lead to convergence may be stronger. We study a model that can capture such effects in \citet{BermanPetersAdamou2019}.

\FloatBarrier

\Fref{ddp_us_fr_trunc} shows that in the United States the ergodicity gap remained stable between 1966 and the early 1980s. Since then it has grown. In France, the ergodicity gap has been roughly stable since 1970. \Fref{top10_MLD} depicts the evolution of the top 10\% income share and the ergodicity gap in the United States and in France over the last 50 years. It demonstrates that both measures evolve similarly in both countries.

\begin{figure}[!htb]
\centering
\includegraphics[width=1.0\textwidth]{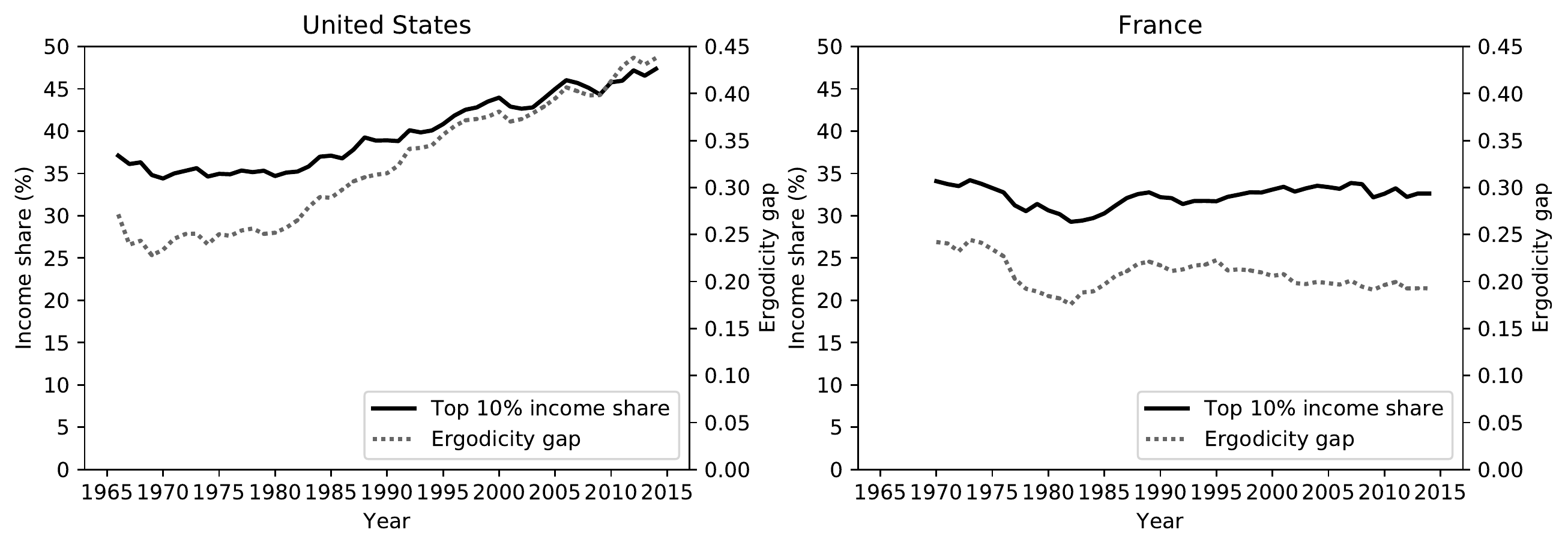}
\caption{The evolution of the top 10\% income share and the ergodicity gap in the United States (left) and in France (right). The correlation between the top 10\% share and the ergodicity gap was 0.94 in the United States and 0.79 in France. All data were taken from \citet{WID2020}.}
\flabel{top10_MLD}
\end{figure}


\section{Discussion}

Economic growth is characterised by two growth rates: the ensemble-average (GDP per capita) growth rate, which is the growth rate of the average income; and the time-average (DDP per capita) growth rate, achieved by each individual in the long run. When incomes follow random multiplicative growth, the ensemble-average (GDP per capita) growth rate is greater than the time-average (DDP per capita) growth rate, and inequality -- measured as the ergodicity gap (MLD) -- increases indefinitely. From the perspective of ergodicity economics, these results are unsurprising: they are direct consequences of the non-ergodicity of the stochastic process for income.

The difference between these growth rates is consequential for policymakers. At present, the most widely used summary statistic of national economic growth is the ensemble-average (GDP per capita) growth rate. This captures growth in the nation's total productive capacity -- referred to colloquially as `the size of the economy' -- and is useful. However, in the model, it is systematically higher than the time-average (DDP per capita) growth rate, which reflects the typical individual's experience of the economy. If these circumstances were mirrored in reality, preoccupation of policymakers with ensemble-average (GDP per capita) growth would create a worrying disconnect between them and the public. There are reasons to believe this situation is present-day reality in the United States, see \fref{ddp_us_fr_trunc} and \citet{BousheyClemens2018,SaezZucman2019b}.

We conclude with some interpretations of these fundamental growth rates. The ensemble-average (GDP per capita) growth rate is the growth rate of average income. By contrast, the time-average (DDP per capita) growth rate is the average growth rate of income. The former is approximately the income-weighted average of individual growth rates and we can think of it, therefore, as a `one dollar, one vote' measure of economic growth (strictly speaking, one dollar per year). The latter is an equally weighted average of individual growth rates and we can think of it as a `one person, one vote' measure. Both have their place in the analysis of economic growth.


\bibliography{../../LML_bibliography/bibliography}

\clearpage

\appendix

\section{Finite populations}\label{app:finite}

In \Secref{theory} we defined the ensemble-average and time-average growth rates of income. By ``ensemble-average growth rate'' we mean the growth rate of the ensemble average of income (not the ensemble average of the growth rate).
In the random multiplicative growth model, this is:
\be
\ga \equiv \frac{\D\ln\ave{x(t)}}{\Dt}\,.
\elabel{ga_app}
\ee
The time-average growth rate is the $\Dt\to\infty$ limit of growth rate of individual income, which we express as the ensemble average of the ergodic growth rate:
\be
\gt \equiv \ave{\frac{\D\ln x(t)}{\Dt}}\,.
\elabel{gt_app}
\ee
These are deterministic quantities, which are properties of the model. They can be expressed in terms of model parameters, as we do for GBM in \secref{example}.

For a finite population of $N$ individuals, we use the finite-$N$ versions of these growth rates:
\be
\gplu = \frac{\D\ln\aveN{x(t)}}{\Dt}; \quad \gdem = \aveN{\frac{\D\ln x(t)}{\Dt}}\,.
\elabel{gr_finN}
\ee
These growth rates are statistics that can be computed from observations of a real population. In the model, they are random variables.

Here we verify that, in a model population of the size of a typical nation, the finite-$N$ growth rates in \eref{gr_finN} are good approximations of their $N\to\infty$ limits. Our aim is to establish whether these limiting growth rates would be relevant in practice, if the incomes of a real finite population were well described by random multiplicative growth. Of course, unmodelled effects -- such as reallocation of income and correlation of income changes -- may be important in reality, and conclusions drawn from the limiting growth rates of a model may be inapt. In our empirical study in \secref{data}, we do not use asymptotic results: rather we estimate finite-$N$ growth rates and watch them evolve over time.

We use GBM, presented in \secref{example} and in greater detail in \appref{model}, as our model of income, where $\ga=\mu$ and $\gt=\mu-\sigma^2/2$. For the ensemble-average growth rate, we use numerical simulations with $\mu=0.02$ per year and $\sigma^2$ ranging from 0 to 0.09 per year.\footnote{Realistically, for wealth, $\sigma^2$ is less than 0.04 per year in the US \citep{BermanPetersAdamou2019}.} We let $N$ change from 2 to $10^7$. For each value of $\sigma$ and $N$, we simulate one time step (equivalent to one year) in a GBM starting from a lognormal distribution, and we compute $\gplu$. We wish to show that $\gplu$ is distributed tightly enough that it is usually closer to $\mu$ than to $\mu-\sigma^2/2$. We do this by repeating the simulation $10^4$ times for each value of $\sigma$ and $N$, creating a sample of $\gplu$ values. We then check what fraction of the sample satisfies $\gplu > \mu -\sigma^2/4$. The left panel in \Fref{gbm_gplu_dem_contour} shows this fraction as a function of $\sigma$ and $N$. It is above 90\% for typical national populations and realistic values of $\sigma$.

\begin{figure}[!htb]
\centering
\includegraphics[width=1.0\textwidth]{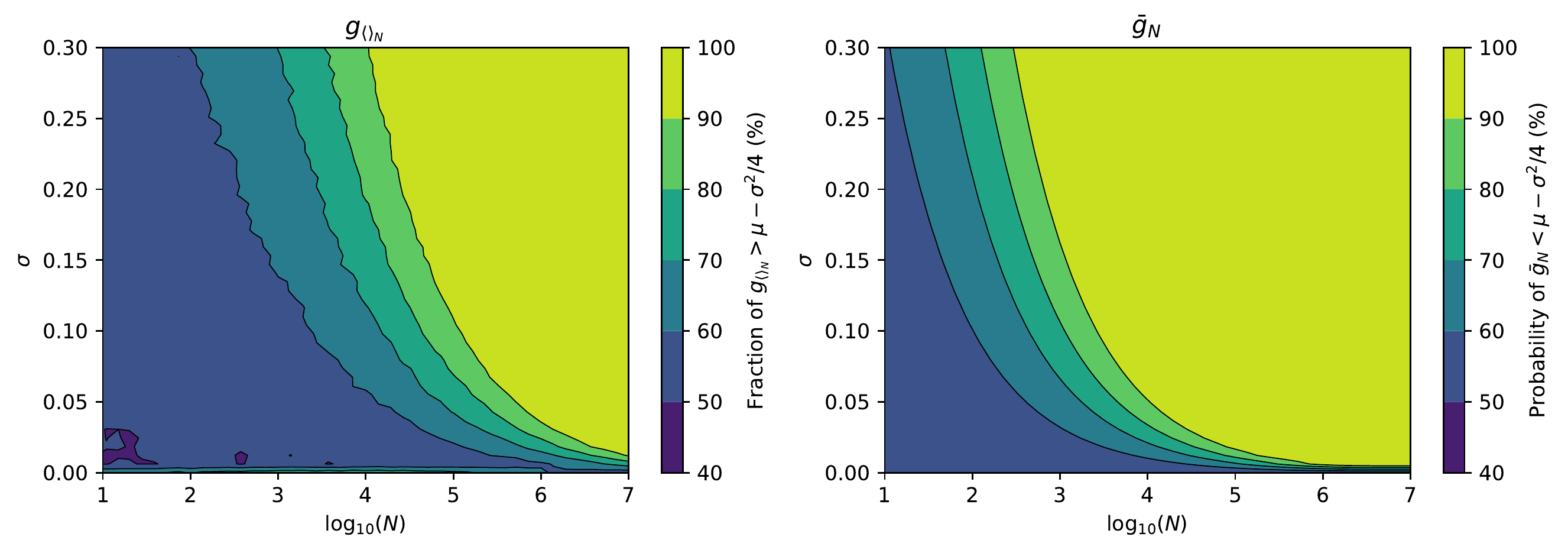}
\caption{Left: The fraction of $\gplu$ values closer to  $\mu$ than to $\mu-\sigma^2/2$ as a function of $\sigma$ and $N$ in $10^4$ realisations of GBM; Right: The probability of $\gdem$ being closer to $\mu-\sigma^2/2$ than to $\mu$ as a function of $\sigma$ and $N$, see \Eref{gdem_tight}.}
\flabel{gbm_gplu_dem_contour}
\end{figure}

We conduct a similar analysis for the time-average growth rate, now requiring that $\gdem$ is usually closer to $\mu-\sigma^2/2$ than to $\mu$. Here the analysis is simpler, since in the model $\gdem$ is just the sample mean of a normally distributed growth rate. Its sampling distribution is
\be
\gdem \sim \N\left(\mu-\frac{\sigma^2}{2},\frac{\sigma^2}{N\Dt}\right)\,,
\ee
and we are interested in the probability $P\left(\gdem < \mu - \sigma^2/4\right)$. For the normal distribution this is
\be
\frac{1}{2} + \frac{1}{2}\erf\left(\sqrt{\frac{\sigma^2 N\Dt}{32}}\right)\,.
\elabel{gdem_tight}
\ee
The right panel in \Fref{gbm_gplu_dem_contour} shows that this probability is above 90\% for typical national populations and realistic values of $\sigma$.

\FloatBarrier

\section{Geometric Brownian motion}\label{app:model}

In \secref{example} we use GBM as an example model of income dynamics. Here we present derivations of the results described therein.

We recall the basic multiplicative model of income evolution in \eref{disc_mult},
\be
x(t+\dt) = x(t)(1+\epsilon)\,,
\elabel{disc_mult_app}
\ee
where $\dt$ is the time between income observations and $\epsilon$ is a random variable. 

In GBM, periodic income multipliers are independent and identically distributed normal random variables. We introduce this using finite-size increments. Income evolves as
\be
x(t+\dt) = x(t)(1+\mu\dt+\sigma\zeta\sqrt{\dt})\,,
\elabel{disc_gbm_1}
\ee
where $\zeta$ is a standard normal variate, $\zeta\sim\N(0,1)$. The model has two parameters, the drift $\mu$ and the volatility $\sigma$. Here $\mu\dt$ is the common deterministic component of the income multiplier, and $\sigma\zeta\sqrt{\dt}$ is the idiosyncratic stochastic part.

In effect, we have chosen $\epsilon\sim\N(\mu\dt,\sigma^2\dt)$ in \eref{disc_mult_app}. This scaling of mean and variance with time ensures that the random variable for one long time step has a distribution consistent with that obtained by summing successive random variables over smaller time steps of equal total duration. This matters because results should not depend on how we partition time.

Defining the change in income at time $t$ as
\be
\dx = x(t+\dt)-x(t)\,,
\ee
we can write \eref{disc_gbm_1} as
\be
\dx = x(t)(\mu\dt + \sigma\zeta\sqrt{\dt})\,.
\elabel{disc_gbm_2}
\ee
We now take the limit $\dt\to0$ to obtain an \Ito stochastic differential equation for income $x(t)$,
\be
dx = x(\mu dt + \sigma dW)\,.
\elabel{gbm_app}
\ee
Thus we arrive at \eref{gbm} in the main text.

\Eref{gbm_app} is solved by using \Ito's formula to find the stochastic differential equation obeyed by $\ln x(t)$. This is a Brownian motion with drift,
\be
d\ln x = \left(\mu - \frac{\sigma^2}{2}\right)dt + \sigma dW\,,
\elabel{bm}
\ee
which is integrated to give the solution for a single trajectory of the stochastic process,
\be
\ln x(t+\Dt) = \ln x(t) + \left(\mu - \frac{\sigma^2}{2}\right)\Dt + \sigma W(\Dt)\,,
\elabel{gbm_sol_1}
\ee
or
\be
x(t+\Dt) = x(t)\exp\left[\left(\mu - \frac{\sigma^2}{2}\right)\Dt + \sigma W(\Dt)\right]\,.
\elabel{gbm_sol_2}
\ee
$\Dt$ is the period of time over which the process runs and $W(\Dt)$ is a Wiener process, with distribution $\N(0,\Dt)$.

\paragraph{Distribution}
Unlike the general multiplicative process in \eref{disc_mult_app}, we see from \eref{gbm_sol_1} that the change in logarithmic income is normally distributed at all times rather than just for long time. The distribution is
\be
\ln x(t+\Dt) - \ln x(t) \sim \N\left(\left(\mu-\frac{\sigma^2}{2}\right)\Dt, \sigma^2\Dt\right)
\elabel{lognormal}
\ee
and it is time-dependent. In particular, its variance grows linearly in $\Dt$.

\paragraph{Exponential growth}
Income grows exponentially. The growth rate,
\be
g = \frac{\D\ln x}{\Dt}\,,
\elabel{g}
\ee
of a single income over the period $\Dt$ is a random variable, which we can extract from \eref{gbm_sol_1} as
\be
g = \mu - \frac{\sigma^2}{2} + \frac{\sigma W(\Dt)}{\Dt}\,.
\ee
This has distribution
\be
g  \sim \N\left(\mu-\frac{\sigma^2}{2}, \frac{\sigma^2}{\Dt}\right)\,.
\elabel{gdist}
\ee

From these results we derive the ensemble-average and time-average growth rates in a GBM, by taking the limits $N\to\infty$ and $\Dt\to\infty$ respectively.

\paragraph{Ensemble-average growth rate}
We recall the definition of $\ga$ in \eref{ga},
\be
\ga \equiv \frac{\D\ln\ave{x(t)}}{\Dt}\,.
\ee
We evaluate this by taking the ensemble average of both sides of \eref{gbm_sol_2} to get
\be
\ave{x(t+\Dt)}=\ave{x(t)}\exp(\mu\Dt)\,,
\ee
the growth rate of which we obtain by inspection as
\be
\ga = \mu\,.
\elabel{ga_gbm}
\ee

\paragraph{Time-average growth rate}
The time-average growth rate is the long-time limit of the individual growth rate in \eref{g}, \ie
\be
\gt \equiv \lim_{\Dt\to\infty}\left\{g\right\}\,.
\ee
The variance of $g$ in \eref{gdist} tends to zero as $\Dt\to\infty$. Therefore, the growth rate converges to the mean of the distribution in this limit, \ie
\be
\gt = \mu-\frac{\sigma^2}{2}\,.
\elabel{gt_equiv}
\ee

\FloatBarrier

\section{Estimating $\gdem$ with quantile data}\label{app:quantile}

The calculation of the time-average (DDP per capita) growth rate in \eref{ggdem} requires us to estimate $\aveN{\ln x}$ at different points in time. In practice we do this using quantile average data. Instead of directly computing
\be
\gdem = \frac{\D\aveN{\ln x}}{\Dt}\,,
\ee
which would be possible if all $N$ incomes were available, we compute
\be
\gdem \approx \frac{\frac{1}{Q}\sum_{q=1}^{Q}\D {\ln x^{(q)}}}{\Dt}\,,
\elabel{approx}
\ee
where $Q$ is the number of quantiles (\eg $Q=100$ for percentiles) and $x^{(q)}$ is the average value of $x$ within the $q^{\text{th}}$ quantile.

Clearly, the finer the division into quantiles is, the more accurate the approximation becomes. To validate the values of $\gdem$ shown in \fref{ddp_us_fr_trunc}, we compare them to values for $\gdem$ obtained using the lower and upper thresholds of each quantile as the quantile average. These give crude upper and lower bounds for $\gdem$. \Fref{ddp_us_fr_lims} shows very little difference between $\gdem$ as calculated in \eref{approx} and these bounds. We conclude that the sensitivity of our results to division into quantiles is low, and that \eref{approx} with percentile averages is a good approximation of the time-average growth rate in practice.

\begin{figure}[!htb]
\centering
\includegraphics[width=1.0\textwidth]{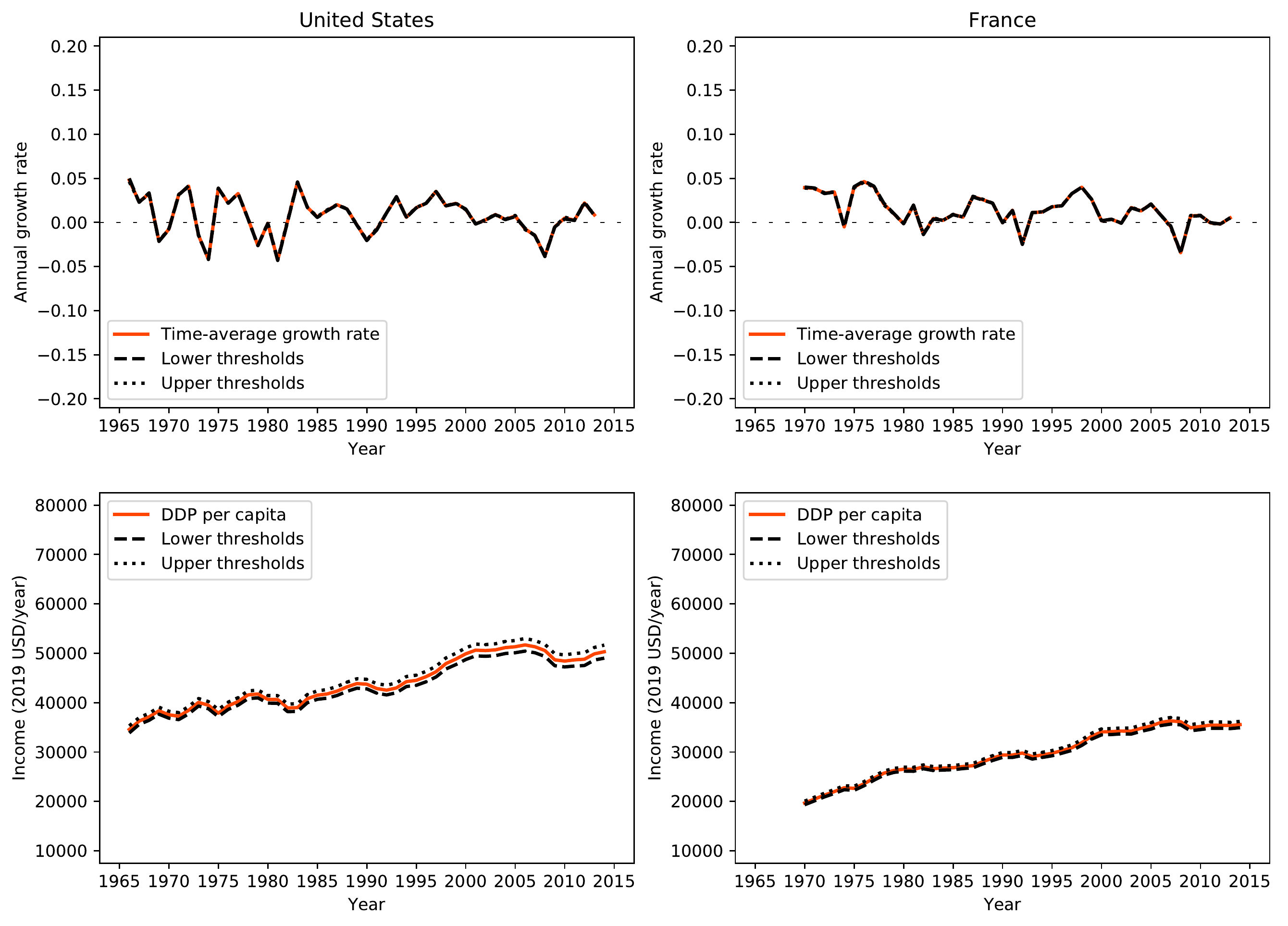}
\caption{The DDP per capita using percentile data. Top: the time-average growth rate (solid red) in the United States (left) and in France (right) estimated using \Eref{approx} based on percentile averages, lower percentile thresholds (dashed black), and upper percentile thresholds (dotted black). Bottom: the evolution of the DDP per capita in the United States (left) and in France (right) based on percentile averages (solid red), lower percentile thresholds (dashed black) and upper percentile thresholds (dotted black). All data were taken from \citet{WID2020}.}
\flabel{ddp_us_fr_lims}
\end{figure}

\FloatBarrier

\section{Truncation of data}\label{app:trunc}

In \secref{theory} we show that DDP per capita can be computed as the geometric mean of income. As such, it requires incomes to be positive and is sensitive to uncertainty in the measurement of low incomes. Therefore, we present results in \fref{ddp_us_fr_trunc} using truncated data, where the bottom 10\% of incomes are excluded from the calculation of growth rates and average incomes.

\Fref{ddp_us_fr_perc} explains why we chose to truncate at this percentile. It shows how the growth of GDP per capita and DDP per capita in the United States and France depend on the fraction of incomes excluded from their computation. Truncating at percentile 1 means that all percentiles are included, while truncating at percentile 100 means that only the top percentile is included (in which case GDP per capita and DDP per capita coincide by construction). In between, DDP per capita growth changes rapidly when the point of truncation varies from percentile 1 to 10, and more slowly after that. This reflects the well-known fact that geometric means are sensitive to changes in small-value data. The arithmetic mean, used to compute GDP per capita, is less sensitive to truncation. Specifically, it shows no major change in truncation-dependence around percentile 10.

\begin{figure}[!htb]
\centering
\includegraphics[width=1.0\textwidth]{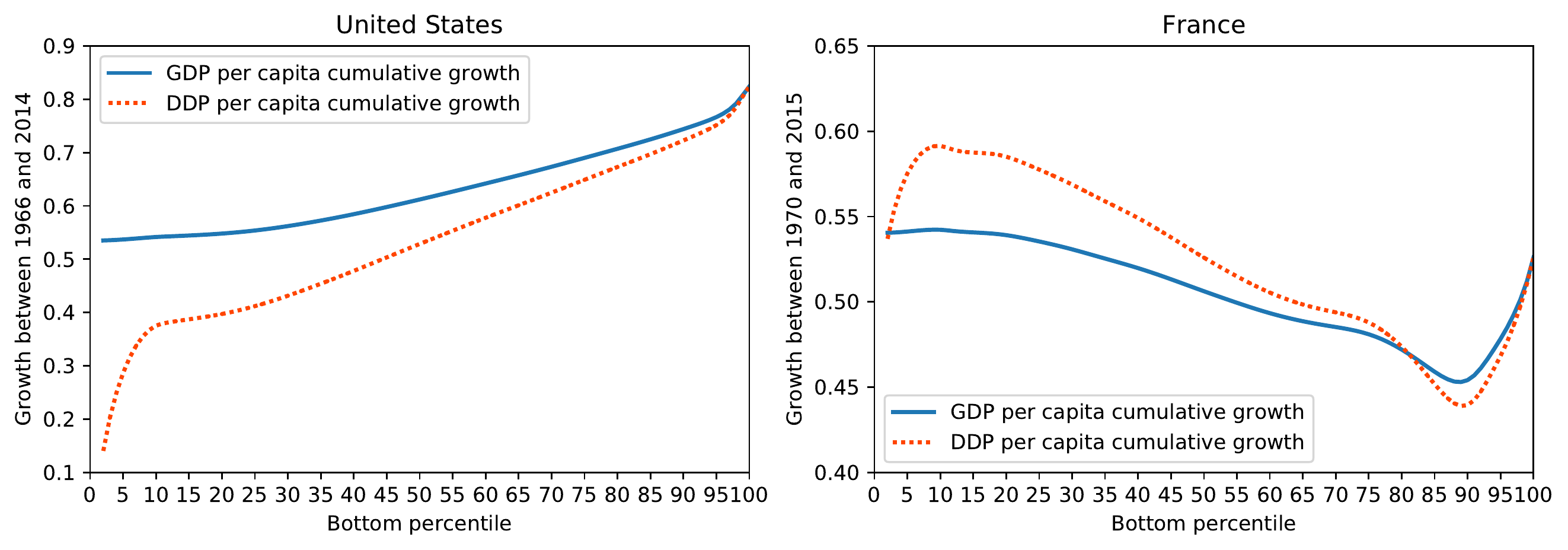}
\caption{The cumulative growth of income depending on point of truncation in the United States and France. The figure depicts the cumulative growth of the GDP per capita (blue) and DDP per capita (red) in the United States between 1966 to 2014 (left), and in France between 1970 to 2015 (right), as a function of the point of truncation. When the truncation is done at percentile 1, all percentiles are included. When it is done at percentile 100, only the top percentile is included and the GDP per capita and DDP per capita coincide. All data were taken from \citet{WID2020}.}
\flabel{ddp_us_fr_perc}
\end{figure}

While truncation has a quantitative effect on the evolution of the GDP and DDP per capita presented in \fref{ddp_us_fr_trunc}, it does not change qualitatively the message. \fref{ddp_us_fr_nontrunc} presents the GDP and DDP per capita and their growth rates in the United States and in France without truncating the distribution. We draw similar conclusions about the differences between the United States and France as we do from the truncated data in \fref{ddp_us_fr_trunc}.

\begin{figure}[!htb]
\centering
\includegraphics[width=1.0\textwidth]{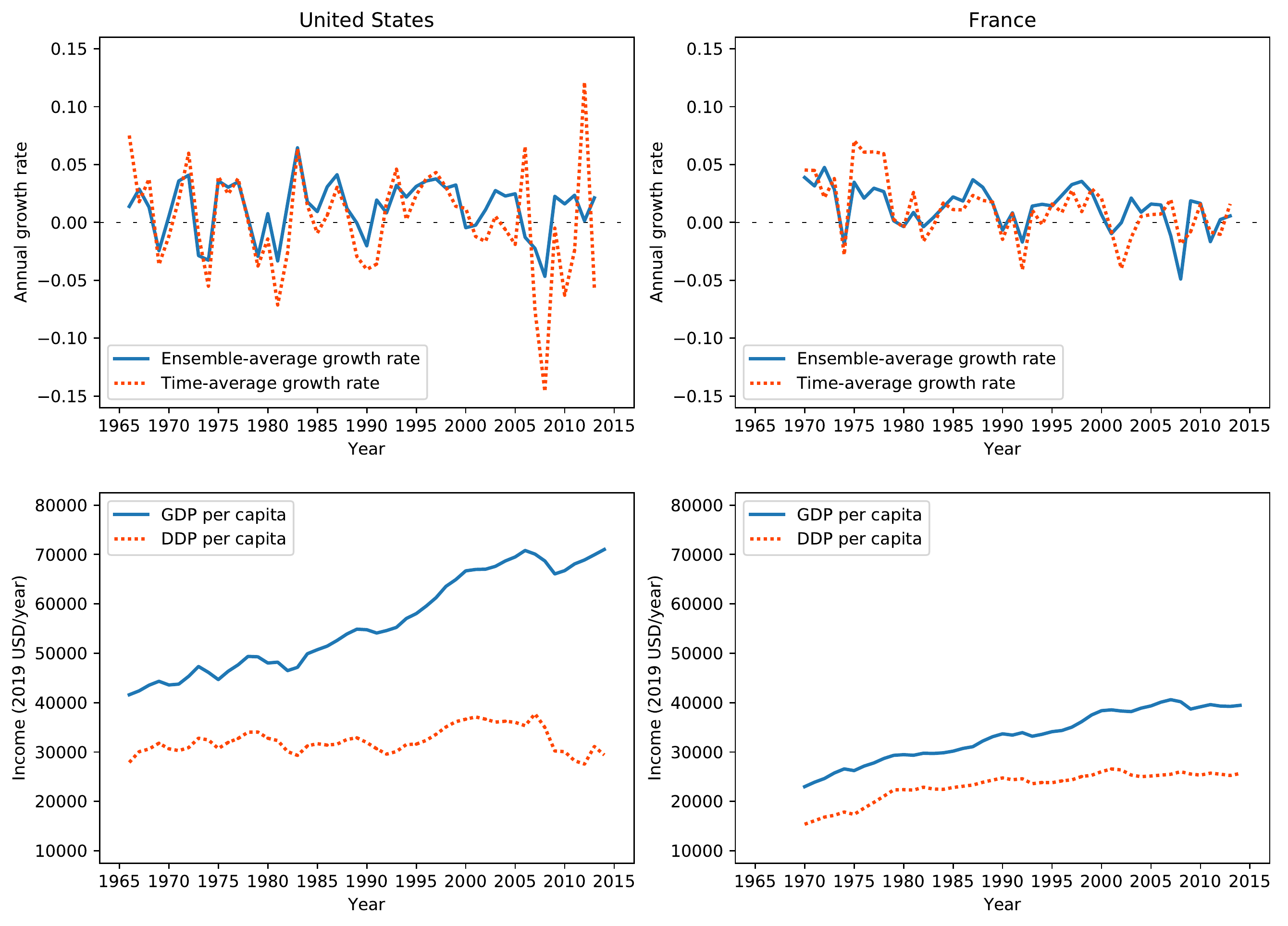}
\caption{The evolution of the two growth rates of the economy. Top: the ensemble-average (blue) and the time-average (red) growth rates in the United States (left) and in France (right). Bottom: the evolution of ensemble-average income (GDP per capita) and DDP per capita in the United States (left) and in France (right). All data were taken from \citet{WID2020}. Compare with \fref{ddp_us_fr_trunc} in which the bottom decile of incomes was excluded from the data in all years.}
\flabel{ddp_us_fr_nontrunc}
\end{figure}

\end{document}